\documentclass[aps,prl,twocolumn,superscriptaddress, floatfix]{revtex4-2}%
\usepackage{times,mathptmx}

\usepackage{float}  

\usepackage{graphicx}
\usepackage{graphics}
\usepackage[caption=false]{subfig}
\usepackage{color}
\usepackage[english]{babel}
\usepackage[utf8]{inputenc}
\usepackage{ae}
\usepackage{latexsym}
\usepackage{dcolumn}
\usepackage{bm}
\usepackage[table]{xcolor}
\usepackage{array}
\usepackage{url}
\usepackage{epsf}
\usepackage{epsfig}
\usepackage{pstricks,pst-grad}
\usepackage[pdftex,colorlinks]{hyperref}
\usepackage{amsmath}
\usepackage{amssymb}
\usepackage{ragged2e}
\usepackage{comment}

\begin{document}

\title[]
{A Variational Approach to Trap Macromolecules}

\author{Luofu Liu}
\affiliation{Department of Chemical and Biomolecular Engineering, University of California Berkeley, Berkeley, California 94720, United States}
\affiliation{Materials Sciences Division, Lawrence Berkeley National Lab, Berkeley, California 94720, United States}

\author{Chao Duan}
\affiliation{Department of Chemical and Biomolecular Engineering, University of California Berkeley, Berkeley, California 94720, United States}

\author{Rui Wang}
\email {ruiwang325@berkeley.edu}\affiliation{Department of Chemical and Biomolecular Engineering, University of California Berkeley, Berkeley, California 94720, United States}
\affiliation{Materials Sciences Division, Lawrence Berkeley National Lab, Berkeley, California 94720, United States}


\begin{abstract}
Trapping macromolecules is impoartant for the study of their conformations, interactions, dynamics and kinetic processes. Here, we develop a variational approach which self-consistently introduces a mean force that controls the center-of-mass position and a self-adjustable harmonic potential that counters the center-of-mass fluctuation. The effectiveness and versatility of our approach is verified in three classical yet not fully understood problems in polymer physics: (1) single-chain conformation in the entire solvent regimes, (2) globule-pearl necklace-coil transition of a polyelectrolyte and (3) inter-chain interaction by simultaneously trapping two polymers. The scaling relationships and $\theta$ behaviors are well captured. Conformations with large shape anisotropy appearing in charged polymers are clearly depicted. Our theoretical predictions are in quantitative agreement with experimental results reported in the literature.
\end{abstract}
\maketitle

Trapping and manipulating individual macromolecules is not only an outstanding challenge in fundamental polymer, soft matter and biophysics, but also an ultimate task in nanotechnology \cite{Maragò2013,Kolbow2021,Bespalova2019,Khan2019,Braun2002,Duhr2006}. It is always desired to counter Brownian motion and manage the position and orientation of macromolecules in a well-controlled manner when placing them in an environment of interest or bringing them into contact with each other \cite{Cohen2005APL,Cohen2005PRL}. The trap of a single molecule enables the probe of its response to external stimuli, e.g. measuring the force-extension of DNA through optical tweezers and detecting protein in nanochannels via electroosmotic flow \cite{Ashkin1987,Smith1992,Wang1997,Bustamante2021,Schmid2021,Hsu2016}. Quantifying the intermolecular interaction requires simultaneous trap of two molecules. It plays a vital role in determining a wealth of structural and dynamic behaviors, such as stability of colloids, solubility of polymers, and affinity of proteins \cite{russel1991, rubinstein2003, Lee1994, Kosuri2019, Wendy2022}. Trapping can also be performed on larger molecular assemblies to uncover key features of complex kinetic processes. The energy barrier and transient state in the kinetic pathway of polyelectrolyte coacervation, protein aggregation, micelle exchange, vesicle fission/fusion, and etc. can thus be identified \cite{Chen2022,Duan2023,Mysona2019PRL,Mysona2019,Lee2008,Zhao2016,Liu2024}.  

The trap of macromolecules has been successfully achieved in experiments by confining them in electromagnetic or flow fields \cite{Bespalova2019,Khan2019,Maragò2013,Kolbow2021,Braun2002,Cohen2005APL,Cohen2005PRL,Duhr2006,Hsu2016}. It has also been realized in simulations via imposing a biased potential \cite{frenkel2023,Withers2003,Yang2019,Chen2022}. However, a robust control of macromolecules through a theoretical approach remains a great challenge. Edward studied the excluded volume effect on the conformation of a single polymer by tethering one chain end, which inevitably introduces a bias and breaks the equivalence of the two ends \cite{Edwards1965}. Based on Lifshitz’s Theory, Grosberg and Kuznetsov confined a polymer via the boundary condition of its density field \cite{Grosberg1992_1,Grosberg1992_3}. However, this method is only effective in poor solvent regime when the polymer forms a compact globule, whereas the trap fails in good solvents due to the large density fluctuation of the coil state \cite{Wang2012,Wang2014}. To enhance the confinement, Xu and Wang introduced a harmonic spring between the middle segment and the center-of-mass (c.m.) in the self-consistent field theory (SCFT) \cite{Xu2021}. Since the spring constant is adopted by assuming the monomer distribution to be Gaussian with spherical symmetry, this approach becomes less accurate away from the $\theta$ regime. The effectiveness of this approach to capture non-spherical conformations or trap two chains simultaneously also has not been verified yet \cite{Dobrynin1996, Minko2002, Duan2020, Liu2022, Liu2024}. Similar problem of presuming the chain conformation also exists in other methods like integral equation approach and renormalized group theory \cite{Krakoviack2004, Schaefer1988,Kruger1989}. To date, there is no unified theory to our knowledge that can quantitatively describe single polymer conformation in the entire solvent regimes, one of the most basic problems in polymer science.

To achieve a robust trap of macromolecules, we develop a variational approach which naturally leads to a force controlling the position of the c.m. and a harmonic potential self-adjusted based on the non-perturbative feedback from the fluctuation of the c.m. To verify the effectiveness and versatility of our approach, we apply it to three classical yet not fully understood problems in polymer physics: (1) single-chain conformation in the entire solvent regimes, (2) globule-pearl necklace-coil transition of a polyelectrolyte, and (3) inter-chain interaction by simultaneously trapping two polymers. Scaling relationships and $\theta$ behaviors are well captured. Our theoretical predictions are in quantitative agreement with experimental results reported in the literature.   

Without loss of generality, we use an example of trapping a single neutral polymer to illustrate the essence of our variational approach. Applications to a charged polymer and the simultaneous trap of multiple chains are shown in Sec. I in the Supplementary Materials. We consider a system of one polymer immersed in $n_{\rm s}$ solvent molecules which are connected with a pure solvent reservoir to maintain the chemical potential of solvent $\mu_{\rm s}$. The polymer is modeled by a continuous Gaussian chain of $N$ Kuhn segments with Kuhn length $b$. To trap this polymer, we fix its c.m. at a specific position $\bm{\xi}$, giving rise to the semicanonical partition function
\begin{equation}\label{eq:partition_function}
    \begin{aligned}
        &Z = \sum_{n_{\rm s}} \frac{e^{\beta \mu_{\rm s} n_{\rm s}}}{n_s !v^{N+n_{\rm s}}} \int {\rm \hat{D}} \{\bm{R}\} \prod_{\gamma = 1}^{n_{\rm s}} \int {\rm d} \bm{r}_{\gamma} \exp(-\beta H)\\
        &\prod_{\bm{r}}\delta[1-\hat{\phi}_{\rm p}(\bm{r}) - \hat{\phi}_{\rm s}(\bm{r})] \delta [\frac{1}{N}\int_0 ^N {\rm d}s \bm{R}(s) - \bm{\xi}]
    \end{aligned}
\end{equation}
where $\int {\rm \hat{D}} \{\bm{R}\}$ denotes the integration over all configurations weighted by Gaussian-chain statistics. The Hamiltonian $\beta H=v^{-1}\chi\int {\rm d} \bm{r} \hat{\phi}_{\rm p}(\bm{r})\hat{\phi}_{\rm s}(\bm{r})$ includes the 
polymer-solvent interaction in terms of the Flory $\chi$ parameter, with $\hat{\phi}_{\rm p}$ and $\hat{\phi}_{\rm s}$ the instantaneous volume fractions of the polymer and solvent, respectively. For simplicity, the volumes of polymer segments and solvent molecules are taken to be the same $v$. The first $\delta$-functional accounts for the incompressibility whereas the second one enforces the constraint on the c.m. of the polymer.

We follow the standard field-theoretical derivations \cite{fredrickson2006} by performing identity transformations to obtain a field representation of the partition function as $Z=Z_0 Z_F$. $Z_0$ denotes the contribution irrelevant to the trap and also exists in other constraint-free systems. $Z_F$ comes from the constraint of the c.m.:
\begin{equation}
Z_F = \int {\rm d} \bm{F} \exp(-L_F)
\end{equation}
where $\bm{F}$ is the force field conjugate to the deviation of c.m. from the targeted position. The ``action" $L_F = -\ln Q$, with $Q$ the single-chain partition function in the auxiliary fields $W_{\rm p}$ (conjugate to polymer density) and $\bm{F}$:
\begin{equation}
\label{eq:single_chain_partition_function}
    Q = \frac{1}{v^{N}}\int {\rm \hat{D}} \{{\bm{R}}\} \exp \bigg\{ -\int_0^N {\rm d}s \bigg[ iW_{\rm p}(\bm{R}(s))
      - i\frac{\bm{F}}{N} \cdot (\bm{R}(s) - \bm{\xi})   \bigg]  \bigg\}
\end{equation}

The lowest-order approximation as used in previous treatment corresponds to taking the saddle-point contribution of $\bm{F}$ \cite{Grosberg1992_3,Liu2022, Liu2024}. It fails to capture the fluctuation of the c.m. which cannot be ignored as the polymer takes a coil state in $\theta$ and good solvents. To self-consistently account for this fluctuation effect, we perform a non-perturbative variational approach using the Gibbs-Bogoliubov-Feynman bound \cite{kardar2007} to evaluate $Z_F$:   
\begin{equation}\label{eq:Feynman_bound}
    Z_{F} = Z_{\rm ref} \langle \exp[-(L_{F}-L_{\rm ref})] \rangle_{\rm ref} \approx Z_{\rm ref} \exp[-\langle L_{F} - L_{\rm ref} \rangle_{\rm ref}]
\end{equation}
where the average $\langle ... \rangle_{\rm ref}$ is taken in the reference ensemble with action $L_{\rm ref}$. $Z_{\rm ref}=\int {\rm d} \bm{F} \exp(-L_{\rm ref})$ is the normalization factor. Here we take $L_{\rm ref}$ to be a $2n$-power modified Gaussian:
\begin{equation}
    e^{-L_{\rm ref}} = \prod_{\kappa = x,y,z} (F_\kappa + if_\kappa)^{2n} \exp \left[ - \frac{(F_\kappa + if_\kappa)^2}{2A_\kappa} \right]
\end{equation}
where the average force $\bm{f}$ and coefficients $A_\kappa$ are taken to be the variational parameters. This construction of $L_{\rm ref}$ assumes the independence of the c.m. fluctuations in the three directions. Note that a general $2n$-power modified Gaussian is adopted here instead of a standard ($0$th-power) Gaussian to reinforce the confinement of the c.m. Although the variational approach using the standard Gaussian \cite{Agrawal2022, Wang2010, Wang2015} can also lead to the correct scaling, this reinforcement is necessary for an accurate prediction of the coil size in $\theta$ and good solvents as we show later.

To focus on the effect of the c.m. fluctuation, we approximate $Z_{F}$ by the right-hand side of Eq. \ref{eq:Feynman_bound}, while taking the saddle-point approximation to evaluate the functional integrals of other fields included in $Z_0$. Minimizing $Z$ with respect to $\bm{f}$, $A_\kappa$ and other fields, we obtained the following constrained self-consistent equations (see Sec. I in the Supplementary Materials) 
\begin{subequations}\label{eq:scf}
\begin{align}
    &w_{\rm p}(\bm{r}) - w_{\rm s}(\bm{r}) = \chi [1-2\phi_{\rm p}(\bm{r})] \label{eq:scf_a}\\
    &\phi_{\rm p}(\bm{r}) = \frac{1}{Q}\int {\rm d}s q(\bm{r},N-s)q(\bm{r},s) \label{eq:scf_b}\\
    & 1- \phi_{\rm p}(\bm{r}) = \exp[\beta \mu_{\rm s} - w_{\rm s}(\bm{r})]  \label{eq:scf_c}\\
    &  \int {\rm d} \bm{r} (\bm{r} - \bm{\xi}) \phi_{\rm p}(\bm{r}) = \bm{0}  \label{eq:scf_d} \\
    & A_\kappa = \lambda / R_{\rm g, \kappa}^2 \label{eq:scf_e}
\end{align}
\end{subequations}
$w_{\rm p}$ and $w_{\rm s}$ are the saddle-point values of the fields conjugate to polymer and solvent densities, respectively. The single-chain partition function $Q = v^{-1} \int {\rm d} \bm{r} q(\bm{r},N-s)q(\bm{r},s)$, where the chain propagator $q$ satisfies:
\begin{equation}\label{eq:propagator}
    \frac{\partial q}{\partial s} = \frac{b^2}{6}\nabla^2 q(\bm{r} ,s)- U_{\rm p}(\bm{r}) q(\bm{r} ,s)
\end{equation}
The total field experienced by the polymer is now: 
\begin{equation}\label{eq:total_field_polymer}
U_{\rm p}(\bm{r}) = w_{\rm p}(\bm{r}) - \frac{1}{N} \bm{f}\cdot (\bm {r} - \bm{\xi}) + \sum_{\kappa = x,y,z}
\frac{\lambda (\kappa-\xi_\kappa)^2}{2N R_{\rm g,\kappa}^2}
\end{equation}
The resulting free energy of the system is
\begin{equation}\label{eq:free_energy}
\begin{aligned}
    &W = -\ln Q  - e^{\beta \mu_{\rm s}}Q_{\rm s} +\lambda \sum_{\kappa = x,y,z} \ln (R_{\rm g, \kappa} / b) \\
    &+ \frac{1}{v} \int{\rm d} \bm{r} \bigg[ \chi \phi_{\rm p} (\bm{r}) \phi_{\rm s}(\bm{r})- w_{\rm p}(\bm{r})\phi_{\rm p}(\bm{r}) - w_{\rm s}(\bm{r})\phi_{\rm s}(\bm{r}) \bigg]
\end{aligned}
\end{equation}

Eq. \ref{eq:total_field_polymer} clearly indicates that the trap is realized via two mechanisms: (1) a mean force $\bm{f}$ to control the c.m. position, and (2) a harmonic potential with the spring constant $\lambda/(N R_{\rm g,\kappa}^2)$ to counter the c.m. fluctuation. $R_{\rm g,\kappa}^2$ is the component of the square radius of gyration in the $\kappa$-direction. $\lambda$ is a function of the power index $n$, which will further be calibrated using the criterion $R_{\rm g}^2 = Nb^2/6$ at $\chi=0.5$ as $N \to \infty$ (see Sec. II in the Supplementary Materials). We want to emphasize that both the force and the harmonic potential are applied to all chain segments, demonstrating the unbiased feature of our approach. The appearance of $R_{\rm g,\kappa}^2$ in the spring constant reflects the non-perturbative feedback of the c.m. fluctuation, enabling a self-adjustable constraint that is only applied on the c.m. without imposing any unnecessary interference on the chain conformation. Differentiating the spring constant in the three directions is necessary to capture conformations beyond spherical symmetry. These are the essential improvements to existing theories. In contrast, a non-variational method using a fixed reference of ideal coil makes the spring constant unchangeable \cite{Xu2021}. The absence of feedback control in such method leads to a wrong scaling of $R_{\rm g}$ in good solvents (see Sec. III in the Supplementary Materials).  

\begin{figure}[t] 
\centering
\includegraphics[width=0.48\textwidth]{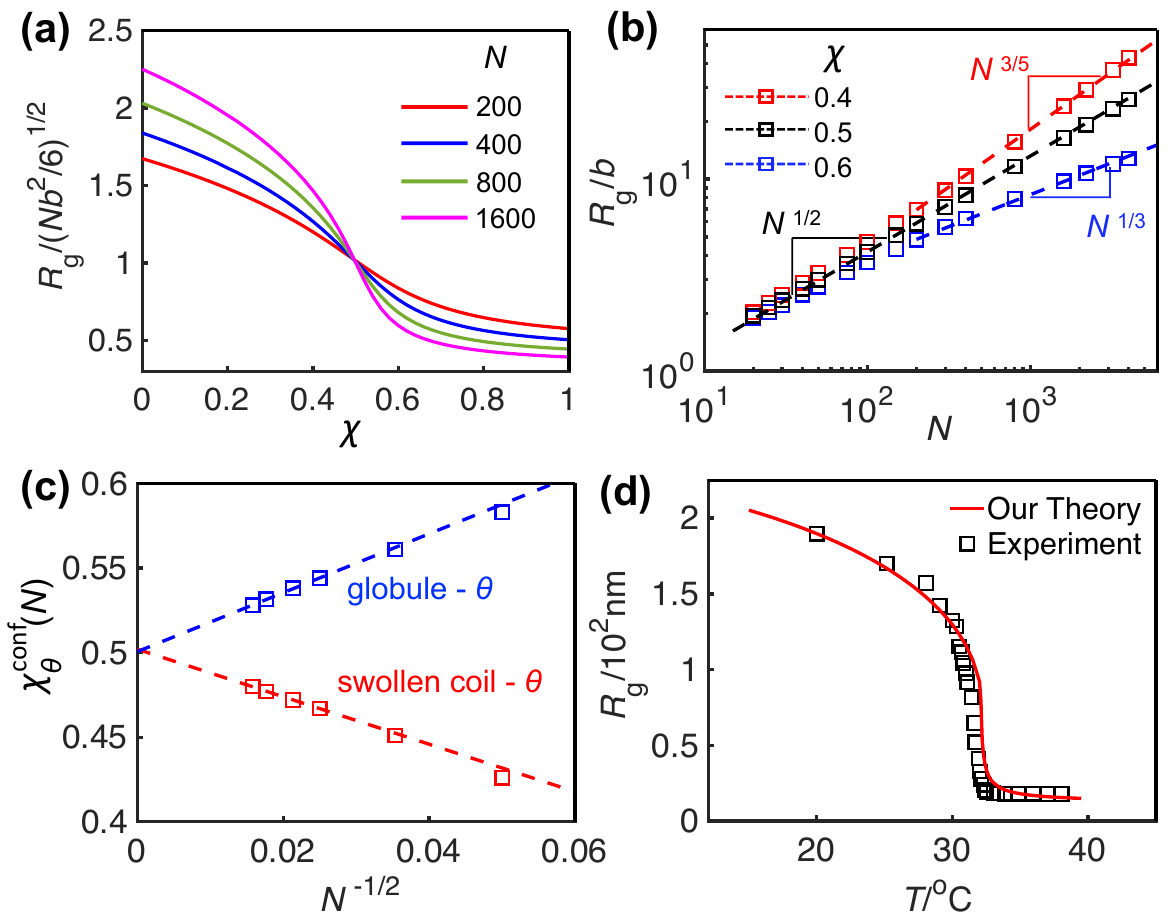}
\caption{Single-chain conformation of a neutral polymer. (a) The radius of gyration $R_{\rm g}$ as a function of $\chi$ for different $N$. (b) The scaling relationship between $R_{\rm g}$ and $N$ in different solvent conditions. (c) The boundary of the $\theta$ regime $\chi_{\theta}^{\rm conf}(N)$ on the globule and coil side, respectively, versus $N^{-1/2}$, where lines represent the linear fit. (d) Comparison of our theoretical prediction with experimental measurements by Wang and Wu \cite{Wu1998} on $R_{\rm g}$ of PNIPAM. $\lambda=1.34$ for the best fit.} 
\label{figure:single-chain}
\end{figure}

\textit{Single-chain conformation of a neutral polymer}.--- We first apply the general theory to trap a single neutral polymer to study its conformation in different solvents. The expected conformations have spherical symmetry, leading to $\bm{f}=\bm{0}$ and the same spring constants in all three directions. Figure \ref{figure:single-chain}a plots the radius of gyration $R_{\rm g}$ as a function of $\chi$ for different chain lengths. As $\chi$ increases, the solvent quality changes from good to poor, reducing polymer size from a swollen coil to a collapsed globule passing the ideal coil state in the $\theta$ regime. The transition becomes sharper as $N$ increases. It is worth noting that the self-adjustable spring constant obtained in our variational approach is the key to effectively counter the c.m fluctuation of the coil and predict the correct density profile and scaling. This is the first time to our knowledge that the single-chain conformation is captured in the entire solvent regimes by a unified theory.

Figure \ref{figure:single-chain}b plots the scaling relationship between $R_{\rm g}$ and $N$. While for $\chi=0.5$, $R_{\rm g}$ shows a single scaling of $N^{1/2}$; for both good ($\chi=0.4$) and poor ($\chi=0.6$) solvents, two scaling regimes exist. Ideal coil scaling of $N^{1/2}$ appears for short chains, which turns to either $N^{3/5}$ (swollen coil) or $N^{1/3}$ (collapsed globule) depending on solvent quality. The turning point signifies the thermal blob size $\xi_{T}$ below which the excluded volume interaction is weaker than the thermal energy $k_{\rm B}T$. $\xi_{T}/b \approx 8$ for $\chi=0.6$. Our results not only confirm the picture of thermal blob but also provide the first identification of its actual size \cite{rubinstein2003}.   

The $\chi$-dependence of $R_{\rm g}$ allows us to probe the $\theta$ regime. The boundary of the $\theta$ regime for finite chain length, $\chi_\theta^{\rm conf}(N)$, is identified by the onset of the rapid size change where $\partial^2R_{\rm g}/\partial \chi^2$ reaches its extremums. Figure \ref{figure:single-chain}c plots $\chi_\theta^{\rm conf}(N)$ on the globule and coil sides of the $\theta$ regime, respectively, both showing a linear relationship with $N^{-1/2}$. The intercepts of the two lines coincide at $\chi_\theta^{\rm conf}(\infty)=0.5$, indicating that the width of the $\theta$ regime reduces to $0$ as $N \to \infty$. The transition becomes discontinuous for infinitely long chain. It also provides a criterion for the calibration of the coefficient $\lambda$ in the spring constant (Eq. \ref{eq:total_field_polymer}) using $R_{\rm g}^2 = Nb^2/6$ at $\chi=0.5$ as $N \to \infty$. $\lambda$ only depends on chain stiffness $b^3/v$, uniquely determined for a given polymer regardless of the chain length. For example, $\lambda=3.61$ for $b^3/v=1$ as used in our calculation.

To further validate our theory, we directly compare our predictions with the experimental measurements by Wang and Wu on the size of poly(N-isopropylacrylamide) (PNIPAM) \cite{Wu1998}. As shown in Fig. \ref{figure:single-chain}d, our theoretical predictions of $R_{\rm g}$ are in quantitative agreement with the experimental data in a wide temperature range. Model parameters $N$, $v$, $b$ and the temperature dependence of $\chi$ are directly taken from the literature (see Sec. IV in the Supplementary Materials) \cite{Wu1998,Nakamoto1996,Ahmed2009}.  

\begin{figure}[t] 
\centering
\includegraphics[width=0.48\textwidth]{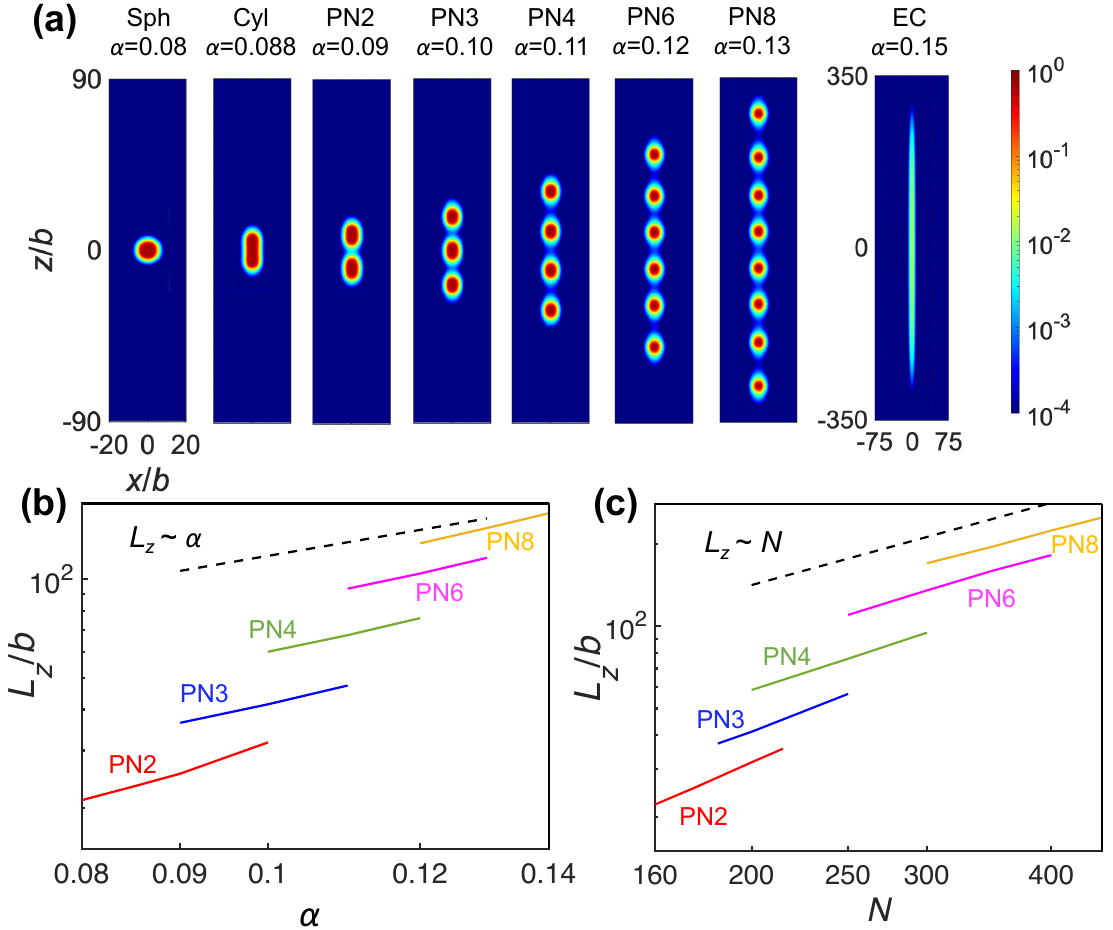}
\caption{Structure of a single polyelectrolyte in a dilute monovalent salt solution. (a) 2D visualizations of the polymer density for the cascade transition from spherical globule (Sph), to cylindrical globule (Cyl), to pearl-necklace with $m$ pearls (PN$m$), and to extended coil (EC). $\lambda=3.61$, $\chi=1.0$, and $N=200$. Dielectric constant $\epsilon=80$ and bulk salt concentration $c_{\rm b}=10^{-4}$M.
The length $L_z$ of PN structure is plotted against (b) the backbone charge density $\alpha$ and (c) chain length $N$, respectively.}
\label{figure:PE}
\end{figure}

\textit{Globule-pearl necklace-coil transition of a polyelectrolyte}.--- It is more challenging to trap a polyelectrolyte chain because of its large c.m. fluctuation and shape anisotropy induced by the Coulomb repulsion between charged monomers \cite{Muthu2023}. Although the mean force $\bm{f}$ remains $\bm{0}$ since the conformation is symmetric with respect to the c.m., the spring constants vary in different directions due to the anisotropic shape change. Figure \ref{figure:PE}a plots the conformation of a single polyelectrolyte with hydrophobic backbone in a dilute salt solution. As the backbone charge density $\alpha$ increases, the polyelectrolyte exhibits a cascade conformational transition in the following sequence: spherical globule $\to$ cylindrical globule $\to$ a series of pearl-necklace with increasing number of pearls $\to$ extended coil. It has been argued whether cylindrical globule or pearl-necklace is a more stable structure, because each has been observed in simulations and experiments \cite{Kiriy2002,Minko2002,Jeon2007,Liao2006,Reddy2006,Spiteri2007,Duan2020}. Our results indicate that both of these two structures can be stable, but in different parameter regimes. In addition, our theory provides a clear depiction of the pearl-necklace structure, able to capture the pearls of the electrostatic-blob size and the necklace of the thermal-blob size at the same time \cite{Dobrynin1996, Dobrynin2005}. We also find that pearl-necklace exhibits odd-even effect: structures with some odd numbers (5 and 7) of pearls are only metastable, whose energy is higher than the adjacent structures with even numbers of pearls. Furthermore, our variational approach remains effective even for structures with extremely high anisotropy, e.g. extended coil with an aspect ratio larger than 100.

Figures \ref{figure:PE}b and \ref{figure:PE}c plot the length $L_z$ of the pearl-necklace against the backbone charge density $\alpha$ and the chain length $N$, respectively. A scaling relationship $L_z \sim \alpha N$ is revealed, where the linearity gets more accurate as the number of pearls increases. Our predictions confirms the result by Dobrynin and Rubinstein using scaling analysis \cite{Dobrynin1996,Dobrynin2005}.

\begin{figure}[t] 
\centering
\includegraphics[width=0.48\textwidth]{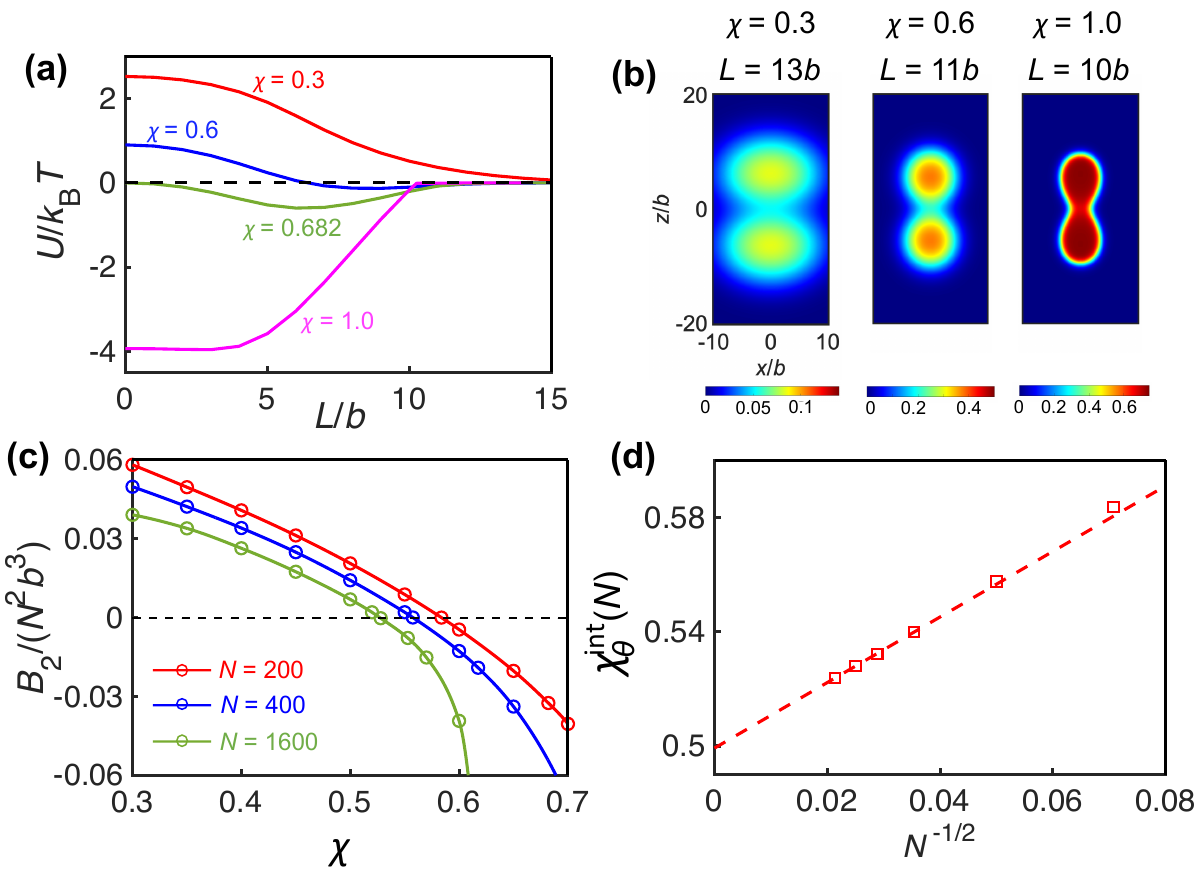}
\caption{Inter-chain interactions between two neutral polymers. (a) Potential of mean force as a function of c.m. separation distance for $N=200$. (b) 2D visualization of the polymer density as two polymers get into contact. (c) Second virial coefficient $B_2$ as a function of $\chi$. (d) $\theta$ point $\chi_{\theta}^{\rm int}(N)$ as a function of $N^{-1/2}$. The dashed line denotes a linear fit.} 
\label{figure:two-chain}
\end{figure}

\textit{Inter-chain interaction between two polymers}.--- The quantification of the interactions between polymers requires a simultaneous trap of multiple chains. For a targeted polymer, the conformational symmetry breaks along the inter-chain axis, which leads to a non-zero mean force $\bm{f}$ in such direction. The anisotropic deformation as polymers approach each other also makes spring constants vary in different directions. We apply our variational approach to trap two neutral polymers at a targeted c.m. separation distance $L$. By tracking the polymer density profile and the associated free energy, we obtain the potential of mean force (PMF) $U(L)$ which is defined by $U(L)=W(L)-W(\infty)$ as the free energy excess to two infinitely separated polymers ($L=\infty$). Figure \ref{figure:two-chain}a plots $U(L)$ for different $\chi$. The inter-chain interaction turns from pure repulsive in good solvent ($\chi=0.3$) to pure attractive in poor solvent ($\chi=1.0$). Near the $\theta$ point, $U(L)$ exhibits long-range attraction and short-range repulsion. This non-monotonic behavior originates from the competition between the monomer two-body attraction and three-body repulsion. The latter becomes dominant at small $L$ where the density of the two polymers largely overlaps, enhancing the excluded volume effect and the resulting three-body repulsion. The non-monotonic shape of PMF predicted by our theory is consistent with the previous simulation results \cite{dautenhahn1994,Krakoviack2003,Withers2003,Pelissetto2005} but cannot be captured by existing theories built upon a fixed Gaussian distribution of chain conformation \cite{Krakoviack2004,Tanaka1985,Raos1996,Grosberg1992_3}. 

The feedback mechanism of our trapping approach enables us to self-consistently capture the evolution of the chain conformation as polymers get into contact. The visualization in Fig. \ref{figure:two-chain}b illustrates how the polymers interfere with each other. The inter-chain repulsion in good solvent compresses the polymers to an oblate shape, similar to squeezing two ``fuzzy balls". In contrast, the inter-chain attraction in poor solvent elongates the polymers and generates a neck between them, analogous to the fusion of two liquid droplets. Very close to $\theta$ point (e.g. $\chi=0.6$ for $N=200$), the shape of each chain is nearly a perfect sphere, almost unaffected by the existence of the other polymer. 

The accurate quantification of the PMF facilitates the evaluation of the second virial coefficient $B_2$, an important quantity that characterizes the effective inter-chain interaction. $B_2$ is calculated by $2\pi \int_0 ^\infty [1 - \exp(-\beta U(L))] L^2{\rm d} L$. Figure \ref{figure:two-chain}c plots $B_2$ as a function of $\chi$, where the $\theta$ point $\chi_\theta^{\rm int}(N)$ is identified at $B_2 = 0$. Note that the previous trapping method using the boundary condition of the polymer density field only works in the deep globule regime such that the determination of the $\theta$ point relies on extrapolation \cite{Liu2022}. In contrast, our variational approach provides the first theoretical quantification of the inter-chain interaction in the entire solvent regimes, enabling a more accurate identification of the $\theta$ point.

Figure \ref{figure:two-chain}d plots $\chi_\theta^{\rm int}(N)$ as a function of $N^{-1/2}$, which shows a good linear relationship consistent with results reported in previous theories and simulations \cite{Withers2003,Zhang2020,Liu2022}. The intercept of the line yields $\chi_\theta^{\rm int}(\infty) = 0.5$ for infinitely long chain. This result coincides with the $\chi_\theta^{\rm conf}(\infty)$ obtained via the single-chain conformation (Fig. \ref{figure:single-chain}c), which highlights the self-consistency of our theory.

In this Letter, we developed a variational approach to trap macromolecules, which self-consistently introduces a mean force that controls the c.m. position of the chain and a harmonic potential resulting from the non-perturbative feedback of the c.m. fluctuation. Our approach has made significant improvements over existing theories in the following aspects. (1) The decoupling of the mean force and the harmonic potential enables the trap of a single chain and multiple chains to be achieved in one framework, which facilitates a unified study of both chain conformation and inter-chain interaction. (2) The mean force and the harmonic potential are applied to all monomers in an unbiased way such that the trap is only performed on the c.m without imposing any unnecessary interference on the chain conformation. (3) The self-adjustable feature of the harmonic potential guarantees a feedback control of the conformational response to a variety of solvent conditions and external stimuli. (4) The differentiation of the spring constants in the three directions allows us to trap conformations with large shape anisotropy. (5) Being a field-theoretical formulation, our trapping approach can be systematically incorporated into the polymeric SCFT as a constraint, which not only enables us to visualize the conformational response and kinetic process, but also facilitates the generalization to macromolecules with various chain architectures, copolymer compositions and charge patterns. 

Our approach provides the first quantification of both the single-chain conformation and inter-chain interaction in the entire solvent regimes. The scaling relationships and $\theta$ behaviors are well captured. We confirm the picture of the thermal blob and first identify its actual size. Applied to charged polymers, our approach effectively captures the full scenario of the globule-pearl necklace-coil transition. Particularly, the two exact different characteristic lengths of the pearl-necklace structure, i.e. electrostatic blob and thermal blob, are clearly depicted. Our theoretical predictions are in quantitative agreement with the experimental results reported in the literature. We verified the effectiveness and versatility of our approach in various systems, from globule to coil, from neutral polymers to polyelectrolytes, and from trapping a single chain to the simultaneous trap of multiple chains. We are now able to put the macromolecules wherever we want, successfully achieving the goal of robust control.

\begin{acknowledgments}
Acknowledgment is made to the donors of the American Chemical Society Petroleum Research Fund for partial support of this research. This research used the computational resources provided by the Kenneth S. Pitzer Center for Theoretical Chemistry.
\end{acknowledgments}

\bibliographystyle{apsrev4-2}
\bibliography{Refs}

\end{document}


\textbf{\huge Supplementary Materials for}\\

\begin{center}
\vspace{0.4cm}
\textbf{\large A Variational Approach to Trap Macromolecules}\\
\vspace{0.4cm}

\vspace{1\baselineskip}
Luofu Liu$^{1,2}$, Chao Duan$^1$, and Rui Wang$^{1,2,*}$

\vspace{2\baselineskip}
1. Department of Chemical and Biomolecular Engineering, University of California Berkeley, CA 94720, USA

\vspace{1\baselineskip}
2. Materials Sciences Division, Lawrence Berkeley National Lab, Berkeley, California 94720, USA

\vspace{1\baselineskip}
* correspondence to: \textcolor{blue}{\underline {ruiwang325@berkeley.edu}}

\end{center}

\renewcommand*{\citenumfont}[1]{S#1}
\renewcommand*{\bibnumfmt}[1]{(S#1)}

\renewcommand{\theequation}{S\arabic{equation}}
\renewcommand{\thefigure}{S\arabic{figure}}

\vspace{3\baselineskip}

\section{I. Derivation of the Variational Approach for Trapping the Center-of-Mass of Polymers}

In this section, we provide a detailed derivation of the variational approach for trapping the center of mass (c.m.) of polymers. The formulation for a single neutral polymer is in given {\bf Sec. 1.1}. The generalization of the approach to a charged polymer is provided in {\bf Sec. 1.2}, and the generalization to the simultaneous trap of two polymers is provided in {\bf Sec. 1.3}.
~\\

\textbf{1.1. Variational Approach for a Single Neutral Polymer}\\

We consider a system of one polymer immersed in $n_{\rm s}$ solvent molecules which are connected with a pure solvent reservoir to maintain the chemical potential of solvent $\mu_{\rm s}$. The polymer is modeled by a continuous Gaussian chain of $N$ Kuhn segments with Kuhn length $b$. To trap this polymer, we fix its c.m. at a specific position $\bm{\xi}$, giving rise to the semicanonical partition function
\begin{equation}\label{eq:partition_function}
    \begin{aligned}
        &Z = \sum_{n_{\rm s}} \frac{e^{\beta \mu_{\rm s} n_{\rm s}}}{n_s !v_{\rm p}^N v_{\rm s}^{n_{\rm s}}}
        \int {\rm \hat{D}} \{\bm{R}\} \prod_{j = 1}^{n_{\rm s}}
        \int {\rm d} \bm{r}_j
        \exp(-\beta H) \prod_{\bm{r}}
        \delta[1-\hat{\phi}_{\rm p}(\bm{r}) - \hat{\phi}_{\rm s}(\bm{r})]
        \delta \left[ \frac{1}{N}\int_0 ^N {\rm d}s \bm{R}(s) - \bm{\xi} \right] 
    \end{aligned}
\end{equation}
where $\int {\rm \hat{D}} \{\bm{R}\}=\int {\rm D} \{\bm{R}\} {\rm exp} [ {-(3/2b^2) \int^N_0 ds (\partial{\bm R}}/{\partial s})^2 ]$ denotes the integration over all configurations weighted by Gaussian-chain statistics. The Hamiltonian
\begin{equation}\label{H1.1}
    \begin{aligned}
        &\beta H=\frac{\chi}{v}\int {\rm d} \bm{r} \hat{\phi}_{\rm p}(\bm{r})\hat{\phi}_{\rm s}(\bm{r})
    \end{aligned}
\end{equation}
 includes the polymer-solvent interaction in terms of the Flory $\chi$ parameter, with $\hat{\phi}_{\rm p}$ and $\hat{\phi}_{\rm s}$ the instantaneous volume fractions of the polymer and solvent, respectively. For simplicity, the volumes of polymer segments and solvent molecules are taken to be the same, $v_{\rm p}=v_{\rm s}=v$. The first $\delta$-functional accounts for the incompressibility whereas the second one enforces the constraint on the c.m. of the polymer.

The transformation from the particle-based to the field-based representation is achieved by inserting the following identities into the partition function:
\begin{align}
1 =& \int {\rm D}\phi_{\kappa} \prod_{\bm r}\delta [ \phi_{\kappa}({\bm r})-\hat{\phi}_{\kappa}({\bm r}) ]
=\int {\rm D}\phi_{\kappa} {\rm D}W_{\kappa} \exp \left\{ i \int d{\bm r} W_{\kappa}({\bm r})
[ \phi_{\kappa}({\bm r})-\hat{\phi}_{\kappa}({\bm r}) ] \right\}  ~~(\kappa={\rm p, s})
\end{align}
where the right-hand side of the equation arises from the Fourier representation of
the $\delta$-function with $W_{\kappa}({\bm r})$ being the Fourier conjugate field to
$\phi_{\kappa}({\bm r})$.
Similarly, the Fourier representations of the incompressibility condition and constraint of the polymer c.m. are
\begin{align}
\delta[1-\hat{\phi}_{\rm p}(\bm{r}) - \hat{\phi}_{\rm s}(\bm{r})]
= \int {\rm D}\Gamma \exp \left\{ i \int {\rm d}{\bm r}\Gamma({\bm r})
\left[1-\hat{\phi}_{\rm p}(\bm{r}) - \hat{\phi}_{\rm s}(\bm{r})\right] \right\}
\end{align}
\begin{align}
\delta \left[ \frac{1}{N}\int_0 ^N {\rm d}s {\bm R}(s) - {\bm \xi} \right] 
=
\int {\rm d} {\bm F} \exp \left\{ i {\bm F} \cdot
\left[ \frac{1}{N}\int_0 ^N {\rm d}s {\bm R}(s) - {\bm \xi} \right]  \right\}
\end{align}
By performing the identity transformations, we obtain the field-based partition function as
\begin{align}\label{SI_eq:field_based_partition_function}
&Z =
\int {\rm D}\phi_{\rm p} {\rm D}\phi_{\rm s} {\rm D}W_{\rm p} {\rm D}W_{\rm s} {\rm D}\Gamma {\rm d}{\bm F}
\sum_{n_{\rm s}} \frac{e^{\beta \mu_{\rm s} n_{\rm s}}}{n_s !v^{N+n_{\rm s}}}
\int {\rm \hat{D}} \{\bm{R}\} \prod_{j = 1}^{n_{\rm s}}
\int {\rm d} \bm{r}_j
\nonumber \\
&\cdot \exp \left\{
\frac{1}{v} \int {\rm d}{\bm r} \left[
-\chi \phi_{\rm p}({\bm r}) \phi_{\rm s}({\bm r})
+iW_{\rm p}({\bm r}) \left( \phi_{\rm p}({\bm r}) - \hat{\phi}_{\rm p}({\bm r}) \right)
+iW_{\rm s}({\bm r}) \left( \phi_{\rm s}({\bm r}) - \hat{\phi}_{\rm s}({\bm r}) \right)
+i\Gamma ({\bm r}) \left( \phi_{\rm p}({\bm r}) + \phi_{\rm s}({\bm r})  - 1 \right) \right] \right\} \nonumber \\
&\cdot \exp \left\{ i {\bm F} \cdot
\left[ \frac{1}{N}\int_0 ^N {\rm d}s {\bm R}(s) - {\bm \xi} \right]  \right\}
\end{align}
Here, Eq. \ref{SI_eq:field_based_partition_function} can be decomposed as $Z=Z_0 Z_F$, where $Z_0$ denotes the contribution irrelevant to the trap and also exists in other constraint-free systems. $Z_F$ comes from the constraint of the c.m. They are given by
\begin{subequations}
\begin{align}
&Z_0 = \int {\rm D}\phi_{\rm p} {\rm D}\phi_{\rm s} {\rm D}W_{\rm p} {\rm D}W_{\rm s} {\rm D}\Gamma 
\exp \left\{
e^{\beta \mu_{\rm s}}Q_{\rm s} +
\frac{1}{v} \int {\rm d}{\bm r} \left[
-\chi \phi_{\rm p} \phi_{\rm s}
+iW_{\rm p} \phi_{\rm p}
+iW_{\rm s} \phi_{\rm s}
+i\Gamma \left( \phi_{\rm p} + \phi_{\rm s} - 1 \right) \right] \right\} \\
&Z_F = \int {\rm d} {\bm F} \exp(-L_F)
\end{align}
\end{subequations}
where $Q_{\rm s}=\int {\rm d}{\bm r} e^{-iW_{\rm s}({\bm r})}$ is the partition function of solvents. The ``action" $L_F = -\ln Q_{\rm P}$, with $Q_{\rm P}$ the single-chain partition function in the auxiliary fields $W_{\rm p}$ and $\bm{F}$:
\begin{equation}
\label{eq:single_chain_partition_function}
    Q_{\rm p} = \frac{1}{v^{N}}\int {\rm \hat{D}} \{{\bm R}\} \exp \bigg\{ -\int_0^N {\rm d}s \bigg[ iW_{\rm p}({\bm R}(s))
      - i\frac{\bm F}{N} \cdot ({\bm R}(s) - {\bm \xi})   \bigg]  \bigg\}
\end{equation}

To focus on the effect of the c.m. fluctuation, we perform a variational treatment on $Z_F$ while taking the saddle-point approximation to evaluate the functional integrals of fields included in $Z_0$ \cite{fredrickson2006}. For the latter case, $W_{\rm p}$, $W_{\rm s}$, and $\Gamma$ will be replaced by their saddle-point values $-iw_{\rm p}$, $-iw_{\rm s}$, and $-i\gamma$. The free energy corresponding to $Z_0$ is given by
\begin{align}\label{eq:W_0}
W_0 &= - \ln Z_0 \nonumber \\
&\approx - e^{\beta \mu_{\rm s}} Q_{\rm s} 
+ \frac{1}{v} \int {\rm d}{\bm r} \left\{
\chi \phi_{\rm p} (\bm{r}) \phi_{\rm s}(\bm{r})- w_{\rm p}(\bm{r})\phi_{\rm p}(\bm{r}) - w_{\rm s}(\bm{r})\phi_{\rm s}(\bm{r})
+\gamma(\bm{r}) \left[ \phi_{\rm p}(\bm{r}) +\phi_{\rm s}(\bm{r}) -1 \right] \right\}
\end{align}
For $Z_F$, the saddle-point contribution of ${\bm F}$, as used in previous work \cite{Grosberg1992_3,Liu2022, Liu2024}, fails to capture the fluctuation of the c.m. which cannot be ignored as the polymer takes a coil state in $\theta$ and good solvents. To self-consistently account for this fluctuation effect, the Gibbs-Feynman-Bogoliubov variational approach is used to estimate the integral of $\bm F$ for evaluating $Z_F$ \cite{kardar2007}:
\begin{equation}\label{eq:Feynman_bound}
    Z_{F} = Z_{\rm ref} \langle \exp[-(L_{F}-L_{\rm ref})] \rangle_{\rm ref} \approx Z_{\rm ref} \exp[-\langle L_{F} - L_{\rm ref} \rangle_{\rm ref}]
\end{equation}
where the average $\langle ... \rangle_{\rm ref}$ is taken in the reference ensemble with action $L_{\rm ref}$. We take the reference action to be the $2n$-power modified Gaussian form centered around the average force $-i{\bm f}$:
\begin{equation}
    e^{-L_{\rm ref}[{\bm F},{\bm f},{\bm A}]} = \prod_{\kappa = x,y,z} (F_\kappa + if_\kappa)^{2n} \exp \left[ - \frac{(F_\kappa + if_\kappa)^2}{2A_\kappa} \right]
\end{equation}
$Z_{\rm ref}=\int {\rm d} \bm{F} \exp(-L_{\rm ref})=\prod_{\kappa}(2\pi)^{1/2}(2n-1)!!A^{n+1/2}_{\kappa}$ is the normalization factor. The average force $\bm{f}$ and coefficients ${A}_\kappa$ are taken to be the variational parameters. This construction of $L_{\rm ref}$ assumes the independence of the c.m. fluctuations in the three directions. Note that a general $2n$-power modified Gaussian is adopted here instead of a standard ($0$th-power) Gaussian to reinforce the confinement of the c.m. Although the variational approach using the standard Gaussian \cite{Agrawal2022, Wang2010, Wang2015} can also lead to the correct scaling (see Fig. \ref{figure:SM_scaling_for_lambda_1}), this reinforcement is necessary for an accurate prediction of the coil size in $\theta$ and good solvents (see Fig. \ref{figure:SM_effect_of_lambda}). Evaluation of $e^{ - \langle L_F \rangle_{\rm ref} }$ gives an approximation of the variational single-chain partition function $Q_{\rm var}$ as follows:
\begin{align}\label{eq:variational_SCPF}
Q_{\rm var} &=
\frac{1}{v^{N}}\int {\rm D} \{{\bm R}\}
\exp \left\{ -\frac{3}{2b^2} \int_0^N {\rm d}s
 \left[ \frac{\partial{\bm R}(s)}{\partial s} \right]^2
-\int_0^N {\rm d}s w_{\rm p}[{\bm R}(s)]
+ {\bm f} \cdot \overline{\Delta{\bm R}} \right\}
\langle  \exp \left( i{\bm g} \cdot \overline{\Delta{\bm R}}  \right)  \rangle_{\rm ref}
\end{align}
where ${\bm g}={\bm F}+i{\bm f}$, and $\overline{\Delta{\bm R}} = N^{-1} \int_0^N {\rm d}s [{\bm R}(s) - {\bm \xi}]$ is the difference of the instantaneous c.m. from the targeted position $\bm{\xi}$. The average in the last term
can be evaluated via the Hubbard-Stratonovich transform \cite{fredrickson2006} as
\begin{align}
\langle  \exp \left( i{\bm g} \cdot \overline{\Delta{\bm R}}  \right)  \rangle_{\rm ref} &=
\frac{1}{Z_{\rm ref}} \prod_{\kappa=x,y,z}
\int {\rm d} g_{\kappa} g^{2n}_{\kappa}
\exp \left( i g_{\kappa} \overline{\Delta R}_{\kappa} - \frac{g^2_{\kappa}}{2A_{\kappa}}  \right)  \nonumber \\
&=\prod_{\kappa=x,y,z} \exp \left( - \frac{A_{\kappa}\overline{\Delta R}^2_{\kappa}}{2} \right)
\sum^n_{m=0} C_{n,m} (A_{\kappa} \overline{\Delta R}^2_{\kappa})^m
\end{align}
with the coefficient $C_{n,m}=(-1)^m[(2n)!(2n-2m-1)!!]/[(2m)!(2n-1)!!]$. The variational
single-chain partition function $Q_{\rm var}$ in Eq. \ref{eq:variational_SCPF} can then be written as
\begin{align}
Q_{\rm var} &=
\frac{1}{v^{N}}\int {\rm D} \{{\bm R}\} e^{-H[\bm R]}
\left[
\prod_{\kappa=x,y,z} \left( \sum^n_{m=0} C_{n,m} A^m_{\kappa} \overline{\Delta R}^{2m}_{\kappa}\right)
\right]
\end{align}
with the Hamiltonian $H[\bm{R}]$ defined by
\begin{align}\label{eq:H_R}
H[{\bm R}] &= \frac{3}{2b^2} \int_0^N {\rm d}s
 \left[ \frac{\partial{\bm R}(s)}{\partial s} \right]^2
+\int_0^N {\rm d}s w_{\rm p}[{\bm R}(s)]
- {\bm f} \cdot \overline{\Delta{\bm R}}
+ \sum_{\kappa=x,y,z} \frac{A_{\kappa}}{2} \overline{\Delta R}^2_{\kappa}
\end{align}
Note that the complete variational single-chain Hamiltonian in $Q_{\rm var}$ should be $H[{\bm R}]-\sum_{\kappa} \ln ( \sum^n_{m=0} C_{n,m} A^m_{\kappa} \overline{\Delta R}^{2m}_{\kappa} )$.
Here, we separate $H[\bm{R}]$ from the complete Hamiltonian to facilitate the evaluation of $Q_{\rm var}$. In this sense, $Q_{\rm var}$ can be re-expressed as an ensemble average based on $H[\bm{R}]$:
\begin{align}
Q_{\rm var} &=
\frac{v^{-N}\int {\rm D} \{{\bm R}\} e^{-H[\bm R]}
\left[
\prod_{\kappa=x,y,z} \left( \sum^n_{m=0} C_{n,m} A^m_{\kappa} \overline{\Delta R}^{2m}_{\kappa}\right)
\right]}{v^{-N}\int {\rm D} \{{\bm R}\} e^{-H[\bm R]}}
\frac{1}{v^N} \int {\rm D} \{{\bm R}\} e^{-H[\bm R]} \nonumber \\
&=\left[ \prod_{\kappa=x,y,z} \left( \sum^n_{m=0} C_{n,m} A^m_{\kappa}
\langle \overline{\Delta R}^{2m}_{\kappa} \rangle_{H}
\right) \right] Q_H
\end{align}
where $Q_H$ is the single-chain partition function with the Hamiltonian $H[\bm R]$
\begin{align}
Q_H=\frac{1}{v^N} \int {\rm D} \{{\bm R}\} e^{-H[\bm R]}
\end{align}
A common way to evaluate $Q_H$ requires replacing the complicated functional integral with the product of chain propagators. The chain propagators satisfy the modified diffusion equation \cite{fredrickson2006}. However, this becomes difficult here due to the nonlinear term appeared in $A_{\kappa}\overline{\Delta R}^2_{\kappa}$ in $H[\bm R]$ (Eq. \ref{eq:H_R}) which contains 2-body contributions.
To circumvent this difficulty, we decompose the Hamiltonian $H[\bm R]$ into a 1-body term $H^{(1)} [{\bm R}(s)]$ and a 2-body term $H^{(2)} [{\bm R}(s),{\bm R}(t)]$:
\begin{align}\label{H1}
H^{(1)}[{\bm R}(s)] &=  \int_0^N {\rm d}s
\left\{
\frac{3}{2b^2} \left[ \frac{\partial{\bm R}(s)}{\partial s} \right]^2
+  w_{\rm p}[{\bm R}(s)]
- \frac{\bm f}{N} \cdot [{\bm R}(s)-{\bm \xi}]
+ \sum_{\kappa=x,y,z} \frac{A_{\kappa}}{2} [R_{\kappa}(s)-\xi_{\kappa}]^2
\right\}
\end{align}
\begin{align}\label{H2}
H^{(2)}[{\bm R}(s),{\bm R}(t)] &=
\int_0^N {\rm d}s \int_0^N {\rm d}^{\prime}t
\left\{ \sum_{\kappa=x,y,z} \frac{A_{\kappa}}{2} [R_{\kappa}(s)-\xi_{\kappa}][R_{\kappa}(t)-\xi_{\kappa}]
\right\}
\end{align}
where the prime in $\int^N_0 {\rm d}^{\prime}t$ indicates the case of $t=s$ is excluded in the integral. Then we can approximate $Q_H$ by evaluating the 2-body part $H^{(2)} [{\bm R}(s),{\bm R}(t)]$ via an averaging under the reference ensemble with 1-body part $H^{(1)} [{\bm R}(s)]$:
\begin{align}\label{H2_from_H1}
Q_H &= \int {\rm D}{\bm R} e^{- H^{(1)}[{\bm R}(s)] - H^{(2)}[{\bm R}(s),{\bm R}(t)]}
= Q^{(1)}_H {\langle e^{- H^{(2)}[{\bm R}(s),{\bm R}(t)]} \rangle}_{H^{(1)}}
\end{align}
with $Q^{(1)}_H=\int D{\bm R} e^{- H^{(1)}[{\bm R}(s)]}$. To evaluate the two-body Hamiltonian in the one-body reference in the right-hand side of Eq. \ref{H2_from_H1}, we first fix the position of the $s$-th monomer ${\bm R}(s)$ and calculate the average with respect to the position of the $t$-th monomer ${\bm R}(t)$ to obtain ${\langle e^{-H^{(2)}[{\bm R}(s),\overline{{\bm R}(t)}] } \rangle}_{H^{(1)}}$. We then take the average with respect to ${\bm R}(s)$ as:
\begin{align}
Q_H &= \int {\rm D}{\bm R} e^{- H^{(1)}[{\bm R}(s)]}
{\langle e^{-H^{(2)}[{\bm R}(s),\overline{{\bm R}(t)}] } \rangle}_{H^{(1)}} \approx \int {\rm D}{\bm R} e^{- H^{(1)}[{\bm R}(s)]-{\langle H^{(2)}[{\bm R}(s),\overline{{\bm R}(t)}] \rangle}_{H^{(1)}} } 
\end{align}
where the second equality assumes that higher order $2n$-body correlations ($n \ge 2$) are neglected. Then $Q_H$ can be evaluated via the forward and backward chain propagators $q({\bm r};s)$, $q^{\dagger}({\bm r};s)$
\begin{align}\label{Q_H_from_q}
Q_H &= \frac{1}{v}
\int d{\bm r} q^{\dagger}({\bm r};s) q({\bm r};s)
\end{align}
\begin{subequations}
\begin{align}\label{q_forward}
& \frac{\partial q({\bm r};s)}{\partial s} =
\frac{b^2}{6} \nabla^2 q({\bm r};s) - U_{\rm p}({\bm r};s) q({\bm r};s)
\end{align}
\begin{align}\label{q_backward}
& - \frac{\partial q^{\dagger}({\bm r};s)}{\partial s} = \frac{b^2}{6} \nabla^2 q^{\dagger}({\bm r};s) - U_{\rm p}({\bm r};s) q^{\dagger}({\bm r};s)
\end{align}
\end{subequations}
where the total interaction field experienced by the $s$-th monomer $U_{\rm p}({\bm r};s)=U^{(1)}_{\rm p}({\bm r})+U^{(2)}_{\rm p}({\bm r};s)$ with the 1-body interaction
\begin{align}\label{U1}
U^{(1)}_{\rm p}({\bm r})=w_{\rm p}({\bm r}) - \frac{\bm f}{N}\cdot ({\bm r}-{\bm \xi})
+ \sum_{\kappa=x,y,z} \frac{A_{\kappa}}{2N} (\kappa -\xi_{\kappa})^2
\end{align}
and the 2-body interaction
\begin{align}\label{U2}
U^{(2)}_{\rm p}({\bm r};s) &= \sum_{\kappa=x,y,z} \frac{A_{\kappa}}{2N^2} (\kappa -\xi_{\kappa})
\Biggl\{
\int^s_0 {\rm d}t
\frac{1}{q_{(1)}({\bm r};s)}
\int {\rm d}{\bm r^{\prime}} (\kappa^{\prime}-\kappa) q_{(1)}({\bm r}^{\prime};t) g_{(1)}({\bm r}^{\prime},{\bm r};t,s)
\nonumber \\
&+ \int^N_s {\rm d}t
\frac{1}{q^{\dagger}_{(1)}({\bm r};s)}
\int {\rm d}{\bm r^{\prime}} (\kappa^{\prime}-\kappa) g_{(1)}({\bm r},{\bm r}^{\prime};s,t) q^{\dagger}_{(1)}({\bm r}^{\prime};t)
\Biggl\}
\end{align}
where the chain propagator $q_{(1)}({\bm r};s)$ ($q^{\dagger}_{(1)}({\bm r};s)$) and intra-chain correlation function $g_{(1)}({\bm r},{\bm r}^{\prime};s,t)$ ($g_{(1)}({\bm r}^{\prime},{\bm r};t,s)$) are subject to the same recurrence relation (Eq. \ref{q_forward} (Eq. \ref{q_backward})) as $q({\bm r};s)$ ($q^{\dagger}({\bm r};s)$) but with interaction field $U^{(1)}_{\rm p}$.

Then the variational free energy corresponding to $Z_F$ is then given by
\begin{align}\label{eq:W_F}
W_F &=-\ln Z_F \nonumber \\
&\approx - \ln Q_H
- \frac{1}{2} \sum_{\kappa=x,y,z} \ln A_{\kappa}
- \sum_{\kappa=x,y,z} \ln \left( \sum^n_{m=0} C_{n,m} A^m_{\kappa} \langle \overline{\Delta R}^{2m}_{\kappa} \rangle_{H} \right)
\end{align}
where the second term in the second equality comes from $-\ln Z_{\rm ref}=-(n+1/2)\sum_{\kappa} \ln A_{\kappa}$ and $-\langle L_{\rm ref} \rangle_{\rm ref}=n\sum_{\kappa} \ln A_{\kappa}$. Note that we ignore the constants independent of the variational parameters.
The last term is from the $2n$-power in the modified Gaussian reference.
Using $\langle x^{2m} \rangle = \int {\rm d}x x^{2m} e^{-ax^2/2+bx} / \int {\rm d}x e^{-ax^2/2+bx}=\sum^m_l (-1)^l (2m-1)!! C_{m,l} b^{2l}a^{-(m+l)}$, the moments $\langle \overline{\Delta R}^{2m}_{\kappa} \rangle_{H}$ can be estimated by neglecting the elastic energy and interaction field $w_{\rm p}$ in Eq. \ref{eq:H_R}.
Thereafter, the last term in Eq. \ref{eq:W_F} becomes a logarithm of linear superposition of power functions for the variational parameter $A_{\kappa}$, in the form of $-\sum_{\kappa} \ln ( \sum^n_{p=0} d_{\kappa,p} A^{-p}_{\kappa} )$. $d_{\kappa,p}$ are prefactors independent of $A_\kappa^{-p}$.

Minimizing $W_F$ with respect to $A_{\kappa}$, we obtain
\begin{align}\label{dW_dA=0}
-\frac{1}{Q_H} \frac{\partial Q_H}{\partial A_{\kappa}}=
 \frac{1}{A_{\kappa}}
\left[ \frac{1}{2}
+ \frac{\sum^n_{p=0} (-p) d_{\kappa,p} A^{-p}_{\kappa}}{\sum^n_{p=0} d_{\kappa,p} A^{-p}_{\kappa}}
\right] ~~(\kappa=x,y,z)
\end{align}
with the help of Eq. \ref{eq:H_R}, the left-hand side yields
\begin{subequations}\label{Rg_square}
\begin{align}
&-\frac{1}{Q_H} \frac{\partial Q_H}{\partial A_{\kappa}}=
\frac{1}{2} \left[ R^2_{g,\kappa} + \delta^{(2)}_{\kappa}  \right] \\
&R^2_{g,\kappa}= \frac{\int {\rm d}{\bm r} (\kappa-\xi_{\kappa})^2 \phi_{\rm p}(\bm r) }{\int {\rm d}{\bm r} \phi_{\rm p}(\bm r)} \\
&\delta^{(2)}_{\kappa}= \frac{1}{\int {\rm d}{\bm r} \phi_{\rm p}(\bm r)} \int {\rm d}{\bm r} (\kappa-\xi_{\kappa}) \phi_{\rm p}(\bm r)
\Biggl\{
\int^s_0 {\rm d}t
\frac{1}{q_{(1)}({\bm r};s)}
\int {\rm d}{\bm r^{\prime}} (\kappa^{\prime}-\kappa) q_{(1)}({\bm r}^{\prime};t) g_{(1)}({\bm r}^{\prime},{\bm r};t,s)
\nonumber \\
&+ \int^N_s {\rm d}t
\frac{1}{q^{\dagger}_{(1)}({\bm r};s)}
\int {\rm d}{\bm r^{\prime}} (\kappa^{\prime}-\kappa) g_{(1)}({\bm r},{\bm r}^{\prime};s,t) q^{\dagger}_{(1)}({\bm r}^{\prime};t)
\Biggl\}
\end{align}
\end{subequations}
where $\delta^{(2)}_{\kappa}$ comes from the 2-body interaction $U^{(2)}_{\rm p}({\bm r};s)$ in Eq. \ref{U2}. For the right-hand side of Eq. \ref{dW_dA=0}, we neglect the $A_\kappa$-dependence of the ratio $r=\sum^n_{p=0} (-p) d_{\kappa,p} A^{-p}_{\kappa}/\sum^n_{p=0} d_{\kappa,p} A^{-p}_{\kappa}$, and treat $r$ as a general fitting parameter for simplicity.
Then Eq. \ref{dW_dA=0} becomes
\begin{align}\label{dW_dA=0_final}
A_{\kappa}=\frac{\lambda}{R^2_{g,\kappa} + \delta^{(2)}_{\kappa}} ~~(\kappa=x,y,z)
\end{align}
with $\lambda=1+2r$.
Similarly, $W_F \approx - \ln Q_H - (\lambda/2) \sum_{\kappa} \ln A_{\kappa}$.
Especially, if we take the standard (0-th power) Gaussian reference, we have $\lambda = 1$.
While the ratio $r$ arising from the $2n$-power modified Gaussian makes $\lambda$ adjustable, so that we can calibrate its value based on the criterion of the known size of the infinitely long ideal chain: $R_{\rm g}^2 = Nb^2/6$ at $\chi=0.5$ as $N \to \infty$.

Taking the saddle-point approximation for the density of polymer and solvents $\phi_{\rm p}(\bm{r})$, $\phi_{\rm s}(\bm{r})$, conjugate fields $w_{\rm p}(\bm{r})$, $w_{\rm s}(\bm{r})$, incompressibility field $\gamma(\bm{r})$, and optimizing $W_F$ with respect to the average force ${\bm f}$, we obtain
\begin{subequations}\label{eq:scf}
\begin{align}
    &w_{\rm p}(\bm{r}) = \chi \phi_{\rm s}(\bm{r}) + \gamma(\bm{r}) \label{eq:scf_a}\\
    &w_{\rm s}(\bm{r}) = \chi \phi_{\rm p}(\bm{r}) + \gamma(\bm{r}) \label{eq:scf_b}\\
    &\phi_{\rm p}(\bm{r}) = \frac{1}{Q_H}\int {\rm d}s q^{\dagger}(\bm{r};s)q(\bm{r};s) \label{eq:scf_c}\\
    & \phi_{\rm s}(\bm{r}) = \exp[\beta \mu_{\rm s} - w_{\rm s}(\bm{r})]  \label{eq:scf_d}\\
    &  \phi_{\rm p}(\bm{r}) + \phi_{\rm s}(\bm{r}) - 1 = 0 \label{eq:scf_e}\\
    &  \int {\rm d} \bm{r} (\bm{r} - \bm{\xi}) \phi_{\rm p}(\bm{r}) = \bm{0}  \label{eq:scf_f}
\end{align}
\end{subequations}
The resulting free energy of the system is
\begin{align}\label{eq:free_energy}
W &= W_0+W_F = -\ln Q_H  - e^{\beta \mu_{\rm s}}Q_{\rm s} - \frac{\lambda}{2} \sum_{\kappa = x,y,z} \ln A_{\kappa} \nonumber \\
&+ \frac{1}{v} \int{\rm d} \bm{r}
\left\{
\chi \phi_{\rm p} (\bm{r}) \phi_{\rm s}(\bm{r})- w_{\rm p}(\bm{r})\phi_{\rm p}(\bm{r}) - w_{\rm s}
(\bm{r})\phi_{\rm s}(\bm{r})
+\gamma(\bm{r}) \left[ \phi_{\rm p}(\bm{r}) +\phi_{\rm s}(\bm{r}) -1 \right]
\right\}
\end{align}
In most cases, we can safely assume that the two-body contribution is negligible compared to the one-body term such that $U^{(2)}_{\rm p}({\bm r};s) \ll U^{(1)}_{\rm p}({\bm r})$ and $\delta_{\kappa}^{(2)} \ll R_{\rm g,\kappa}^2$. Under this assumption, Eqs. \ref{dW_dA=0_final}, \ref{q_forward}, \ref{U1}, and \ref{eq:free_energy} become Eqs. 6e-9 in the main text.

~\\

\textbf{1.2. Generalization to a Charged Polymer}\\

We consider a semicanonical ensemble consisting of one polyelectrolyte (PE) chain connected to a reservoir of a salt solution with bulk ion concentration $c^b_{\pm}$ that maintains the chemical potentials of solvent $\mu_{s}$ and ions $\mu_{\pm}$.
The PE is assumed to be a homogeneous Gaussian chain with smeared backbone charge density $\alpha$ \cite{Duan2020}. Mobile ions are taken as point charges with valency $z_{\pm}$.
The semicanonical partition function can be written as
\begin{align}\label{Z1.2}
Z = \frac{1}{v_{\rm p}^N}
\int {\rm \hat{D}} \{\bm{R}\}
\prod_{\zeta={\rm s},\pm} \sum_{n_{\rm s}} \frac{e^{\beta \mu_{\zeta} n_{\zeta}}}{n_{\zeta} ! v_{\zeta}^{n_{\zeta}}} \prod_{j = 1}^{n_{\zeta}}
\int {\rm d} \bm{r}_j
\exp(-\beta H) \prod_{\bm{r}}
\delta[1-\hat{\phi}_{\rm p}(\bm{r}) - \hat{\phi}_{\rm s}(\bm{r})]
\delta \left[ \frac{1}{N}\int_0 ^N {\rm d}s \bm{R}(s) - \bm{\xi} \right] 
\end{align}
with the Hamiltonian
\begin{align}\label{H1.2}
H= \frac{\chi}{v}\int {\rm d} \bm{r} \hat{\phi}_{\rm p}(\bm{r})\hat{\phi}_{\rm s}(\bm{r})
+\frac{1}{2} \int {\rm d}{\bm r} {\rm d}{\bm r}^{\prime}
\hat{\rho}_e({\bm r}) C({\bm r},{\bm r}^{\prime}) \hat{\rho}_e({\bm r}^{\prime})
\end{align}
where $\hat{\rho}_e({\bm r})=z_+\hat{\rho}_+({\bm r})+z_-\hat{\rho}_-({\bm r})+\alpha \hat{\phi}_{\rm p}({\bm r})/v$ is the net charge density.
Note that $z_{\pm}$ and $\alpha$ contain the sign of their charges, and we have assumed $v_P=v_S=v$.
$C({\bm r},{\bm r}^{\prime})$ is the Coulomb operator which satisfies
\begin{align}
-\nabla \cdot [\epsilon({\bm r})\nabla C({\bm r},{\bm r}^{\prime}) ] =\delta({\bm r}-{\bm r}^{\prime})
\end{align}
where $\epsilon({\bm r})=kT \epsilon_0 \epsilon_r({\bm r}) /e^2$ is the scaled permittivity with $\epsilon_0$ the vacuum permittivity and
$\epsilon_r({\bm r})$ the local dielectric constant which depends on the local composition of the system \cite{Duan2020}.

We follow a similar procedure as \textbf{Sec. 1.1}, which involves:
(1) identity, Fourier and Hubbard-Stratonovich transformations to decouple the interacting system into noninteracting polymer $\phi_{\rm p}({\bm r})$, solvents $\phi_{\rm s}({\bm r})$, and ions in the fluctuating fields $W_{\rm p}({\bm r})$, $W_{\rm s}({\bm r})$, $\Gamma ({\bm r})$, $\Psi ({\bm r})$ and force ${\bm F}$. $\Psi ({\bm r})$ is the electrostatic potential field conjugate to the net charge;
(2) the saddle-point approximation to simplify the evaluation of the functional integral over the fluctuating fields, i. e., $(W_{\rm p}, W_{\rm s}, \Gamma, \Psi) \rightarrow (-iw_{\rm p}, -iw_{\rm s}, -i\gamma, -i\psi)$;
and (3) a variational treatment to account for the fluctuation of force that controls the c.m. position of polymer chain.
Then similar self-consistent field equations as those in Eqs. \ref{eq:scf} can be obtained with the electrostatics relevant modification to the conjugate field of polymer
\begin{align}
w_{\rm p}(\bm{r}) = \chi \phi_{\rm s}(\bm{r}) + \gamma(\bm{r})
-\frac{v}{2}\frac{\partial \epsilon({\bm r}) }{\partial \phi_{\rm p}({\bm r})}
[\nabla\psi({\bm r})]^2 + \alpha\psi({\bm r})
\end{align}
where the electric potential $\psi$ satisfies the Poisson equation:
\begin{align}
-\nabla\cdot[\epsilon({\bm r})\nabla\psi({\bm r})] =
z_+ c_+({\bm r}) + z_- c_-({\bm r})
+\frac{\alpha}{v_0}\phi_{\rm p}({\bm r})
\end{align}
with ion concentration
\begin{align}
c_{\pm}({\bm r}) =\lambda_{\pm} e^{- z_{\pm} \psi({\bm r})}
\end{align}
where $\lambda_{\pm}=e^{\beta \mu_{\pm}}/v_K$ is the fugacity of the salt ion.

Finally, the resulting free energy of the system becomes
\begin{align}\label{eq:free_energy1.2}
W &= -\ln Q_H  - e^{\beta \mu_{\rm s}}Q_{\rm s} - \frac{\lambda}{2} \sum_{\kappa = x,y,z} \ln A_{\kappa} \nonumber \\
&+ \frac{1}{v} \int{\rm d} \bm{r}
\left\{
\chi \phi_{\rm p} (\bm{r}) \phi_{\rm s}(\bm{r})- w_{\rm p}(\bm{r})\phi_{\rm p}(\bm{r}) - w_{\rm s}
(\bm{r})\phi_{\rm s}(\bm{r})
+\gamma(\bm{r}) \left[ \phi_{\rm p}(\bm{r}) +\phi_{\rm s}(\bm{r}) -1 \right]
\right\} \nonumber \\
&+\int {\rm d}{\bm r} \left\{ \frac{\alpha}{v}\phi_{\rm p}({\bm r})\psi({\bm r}) - \frac{\epsilon({\bm r})}{2}[\nabla\psi({\bm r})]^2 -c_+({\bm r}) -c_-({\bm r})  \right\}
\end{align}

~\\

\textbf{1.3. Generalization to the Simultaneous Trap of Two Polymers}\\

Generalization of the variational approach for trapping two polymers is straightforward by independently fixing the c.m. positions of the two indivdidual polymers. For simplicity, without loss of generality, we consider two identical neutral polymers with their c.m. fixed at position ${\bm \xi_1}$ and ${\bm \xi_2}$, respectively.
The semicanonical partition function becomes
\begin{align}\label{eq:partition_function1.3}
Z = \prod_{\zeta=1,2}\frac{1}{v_{\rm p}^N}
\int {\rm \hat{D}} \{\bm R_{\zeta} \} 
\delta \left[ \frac{1}{N}\int_0 ^N {\rm d}s \bm{R}_{\zeta}(s) - {\bm \xi}_{\zeta} \right]
\prod_{j = 1}^{n_{\rm s}}
\sum_{n_{\rm s}} \frac{e^{\beta \mu_{\rm s} n_{\rm s}}}{n_s ! v_{\rm s}^{n_{\rm s}}}
\int {\rm d} \bm{r}_j
\exp(-\beta H) \prod_{\bm{r}}
\delta[1-\hat{\phi}_{\rm p}(\bm{r}) - \hat{\phi}_{\rm s}(\bm{r})]
\end{align}
Following the same procedure as in \textbf{Sec. 1.1}, the polymer density becomes
\begin{align}
\phi_{\rm p}(\bm{r}) &=
\sum_{\zeta=1,2} \phi_{\zeta}(\bm r)
=\sum_{\zeta=1,2}\frac{1}{Q_{\zeta}}\int {\rm d}s q^{\dagger}_{\zeta}(\bm{r};s)q_{\zeta}(\bm{r};s)
\end{align}
where propagators $q_{\zeta}({\bm r};s)$ and $q^{\dagger}_{\zeta}({\bm r};s)$ are subject to the same recurrence relation in Eq. \ref{q_forward} and Eq. \ref{q_backward} as $q({\bm r};s)$ ($q^{\dagger}({\bm r};s)$) with the average force ${\bm f}_{\zeta}$, variational parameter ${\bm A}_{\zeta}$ and c.m. position ${\bm \xi}_{\zeta}$. Similarly, $Q_{\zeta}$ and ${\bm A}_{\zeta}$ can be evaluated the same as $Q_H$ in Eq. \ref{Q_H_from_q} and ${\bm A}$ in Eq. \ref{dW_dA=0_final}, respectively. ${\bm f}_{\zeta}$ can be determined from
\begin{align}
\int {\rm d} {\bm r} ({\bm r} - {\bm \xi}_{\zeta}) \phi_{\zeta}(\bm r) = {\bm 0}
\end{align}

Finally, the semi-canonical free energy is given by
\begin{align}\label{eq:free_energy1.3}
W &=-\sum_{\zeta=1,2} \left( \ln Q_{\zeta}
+\frac{\lambda}{2} \sum_{\kappa = x,y,z} \ln A_{\zeta,\kappa} \nonumber
\right)
- e^{\beta \mu_{\rm s}}Q_{\rm s} \\
&+ \frac{1}{v} \int{\rm d} \bm{r}
\left\{
\chi \phi_{\rm p} (\bm{r}) \phi_{\rm s}(\bm{r})- w_{\rm p}(\bm{r})\phi_{\rm p}(\bm{r}) - w_{\rm s}
(\bm{r})\phi_{\rm s}(\bm{r})
+\gamma(\bm{r}) \left[ \phi_{\rm p}(\bm{r}) +\phi_{\rm s}(\bm{r}) -1 \right]
\right\}
\end{align}

\newpage

\section{II. The Calibration of $\lambda$}
\begin{figure}[h] 
\centering
\includegraphics[width=0.8\textwidth]{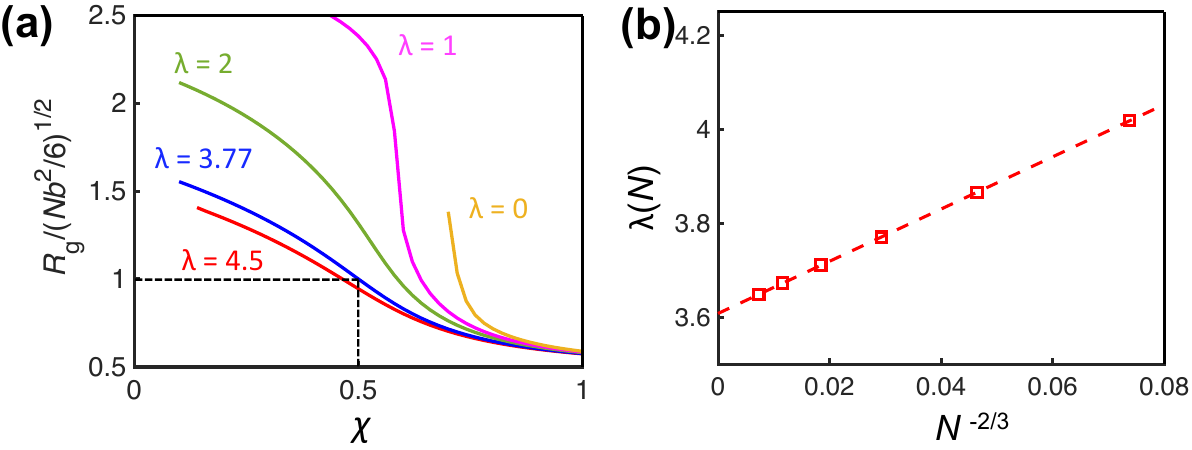}
\caption{Calibration of $\lambda$. (a) $R_{\rm g}$ plotted as a function of $\chi$ for different $\lambda$. $N=200$ and $b^3/v=1$. (b) The plot of $\lambda(N)$ satisfying $R_{\rm g}=\sqrt{Nb^2/6}$ at $\chi=0.5$ for finite $N$. The limiting value yields $\lambda(\infty)=3.61$.} 
\label{figure:SM_effect_of_lambda}
\end{figure}

In this Section, we present the details of the calibration of $\lambda$. Figure \ref{figure:SM_effect_of_lambda}a plots $R_{\rm g}$ as a function of $\chi$ for different $\lambda$. The reference action $L_{\rm ref}$ using a standard Gaussian corresponds to $\lambda = 1$, which largely overestimates $R_{\rm g}$ at $\chi=0.5$ as $R_{\rm g} \approx 2.5 \sqrt{Nb^2/6}$. This deviation indicates that it is necessary to calibrate $\lambda$ for a quantitative prediction of $R_{\rm g}$. $R_{\rm g}$ decreases as $\lambda$ increases. For the case of $N=200$ and $b^3/v=1$, $R_{\rm g}$ exactly equals $\sqrt{Nb^2/6}$ at $\chi=0.5$ when $\lambda = 3.77$. Further increasing $\lambda$ leads to an underestimation of $R_{\rm g}$. Therefore, the first step of the calibration is to determine the $\lambda$ value at a finite chain length $N$, $\lambda(N)$, which satisfies $R_{\rm g}=\sqrt{Nb^2/6}$ at $\chi=0.5$.

The next step is to extrapolate $\lambda(N)$ to $N\to\infty$. As plotted in Fig. \ref{figure:SM_effect_of_lambda}b, $\lambda(N)$ has a linear relationship with $N^{-2/3}$: 
\begin{equation}
    \lambda(N) = \lambda(\infty) + \alpha N^{-2/3}
\end{equation}
The intercepts tells $\lambda(\infty)$ which satisfies the well-known criterion $R_{\rm g}^2 = Nb^2/6$ at $\chi=0.5$ as $N \to \infty$. $\lambda$ only depends on chain stiffness $b^3/v$, uniquely determined for a given polymer regardless of the chain length. For example, $\lambda=3.61$ for $b^3/v=1$ as used in our calculation.

\begin{figure}[h] 
\centering
\includegraphics[width=0.5\textwidth]{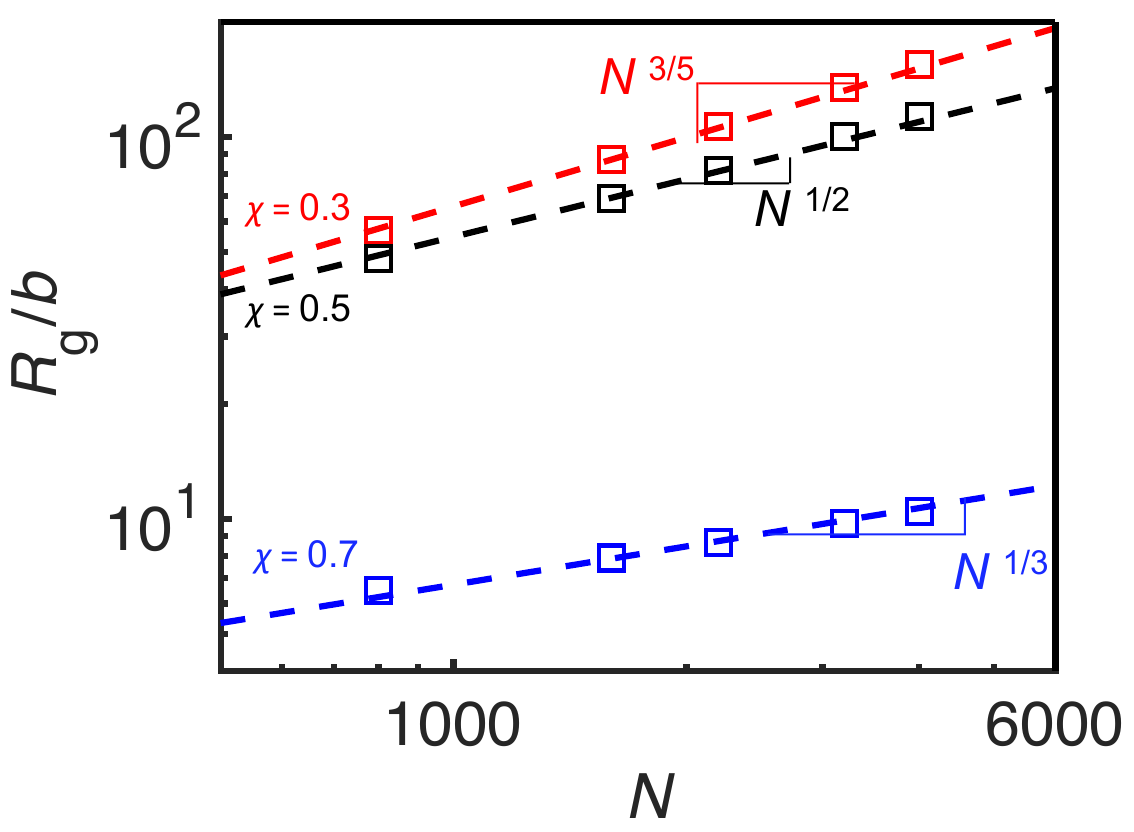}
\caption{Scaling relationship between $R_{\rm g}$ and $N$ in good, $\theta$ and poor solvents for $\lambda=1$.} 
\label{figure:SM_scaling_for_lambda_1}
\end{figure}

It should be noted that, while taking $L_{\rm ref}$ to be a standard Gaussian (i.e. $\lambda=1$) fails to accurately predict $R_{\rm g}$, it can still predict the correct scaling relationship between $R_{\rm g}$ and $N$ in good, $\theta$ and poor solvent regimes, as shown in Fig. \ref{figure:SM_scaling_for_lambda_1}.

\newpage
\section{III. Comparison with the non-variational method using a fixed reference of ideal coil}
\begin{figure}[h] 
\centering
\includegraphics[width=0.6\textwidth]{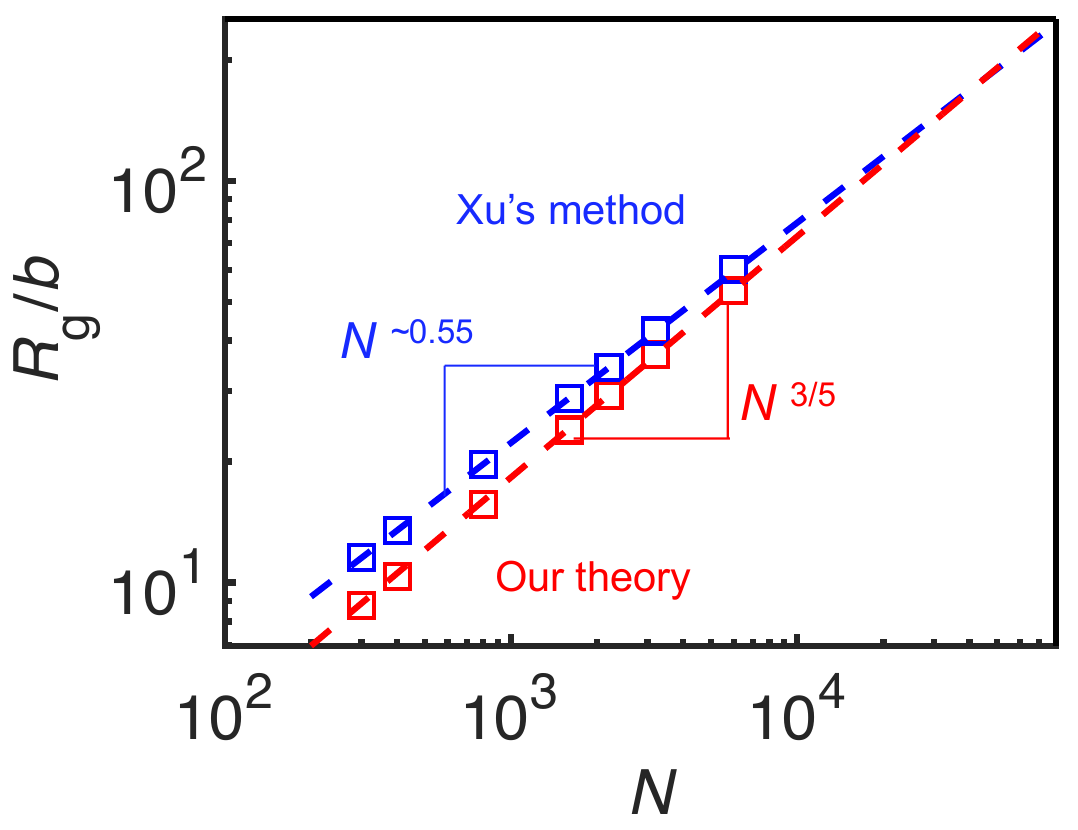}
\caption{Comparison between our variational approach and Xu's method of constraining the middle segment via a harmonic spring \cite{Xu2021}. The radius of gyration $R_{\rm g}$ is plotted against $N$ for both methods. $b^3/v=1$. $\lambda=3.61$ is used in our theory.} 
\label{figure:compare_with_Xu}
\end{figure}

In this section, we compare our variational approach with Xu's method of constraining the middle segment via a harmonic spring \cite{Xu2021}. As shown in Fig. \ref{figure:compare_with_Xu}, our variational approach predicts the $N^{3/5}$ scaling, consistent with the well-known result in the absence of density fluctuation. In contrast, the application of Xu's method to polymers in good solvent yields a wrong scaling as $R_{\rm g} \sim N^{\sim 0.55}$. In Xu's method, the spring constant is derived from a fixed reference of the ideal coil conformation (where $R_{\rm g} \sim N^{1/2}$), which fails to capture the feedback of the actual polymer conformation. The application of this spring constant to polymers in the swollen coil conformation leads to such deviation in the scaling exponent.

\newpage
\section{IV. Parameterization for Comparison in Fig. 1d}
In this Section, we provide the details of the parameterization for the comparison with experiments in Fig. 1d in the main text. Ahmed et al. studied the kinetics of PNIPAM hydrophobic collapsing with a UV resonance Raman spectroscopy  \cite{Ahmed2009}. Their results indicate that one Kuhn monomer of PNIPAM contains $\approx 10$ chemical monomers, leading to the molecular weight of a Kuhn monomer $M_k = 1131.6{\rm g/mol}$. Using the single-chain molecular weight of $M=1.3*10^7 {\rm g/mol}$ in Wu and Wang's work \cite{Wu1998}, the chain length in terms of the number of Kuhn monomers is given by $N=M/M_k\approx11488$. The volume of a Kuhn segment $v$ can be calculated from its molecular weight and density ($\rho = 1.1{\rm g/cm^3}$) as $v=M_k/\rho\approx 1.7 \rm{nm^3}$.

We calculate the Kuhn length $b$ from the radius of gyration of PNIPAM at its ideal coil state, $R_{\rm g, id}$. This state is estimated to be the position where $R_{\rm g}$ has the most rapid change with temperature. $R_{\rm g, id}$ is estimated to be $\approx 80 {\rm nm}$ from Wu and Wang's work \cite{Wu1998}, giving the Kuhn length $b=R_{\rm g, id}/\sqrt{N/6}\approx 1.8{\rm nm}$.

\begin{figure}[h] 
\centering
\includegraphics[width=0.6\textwidth]{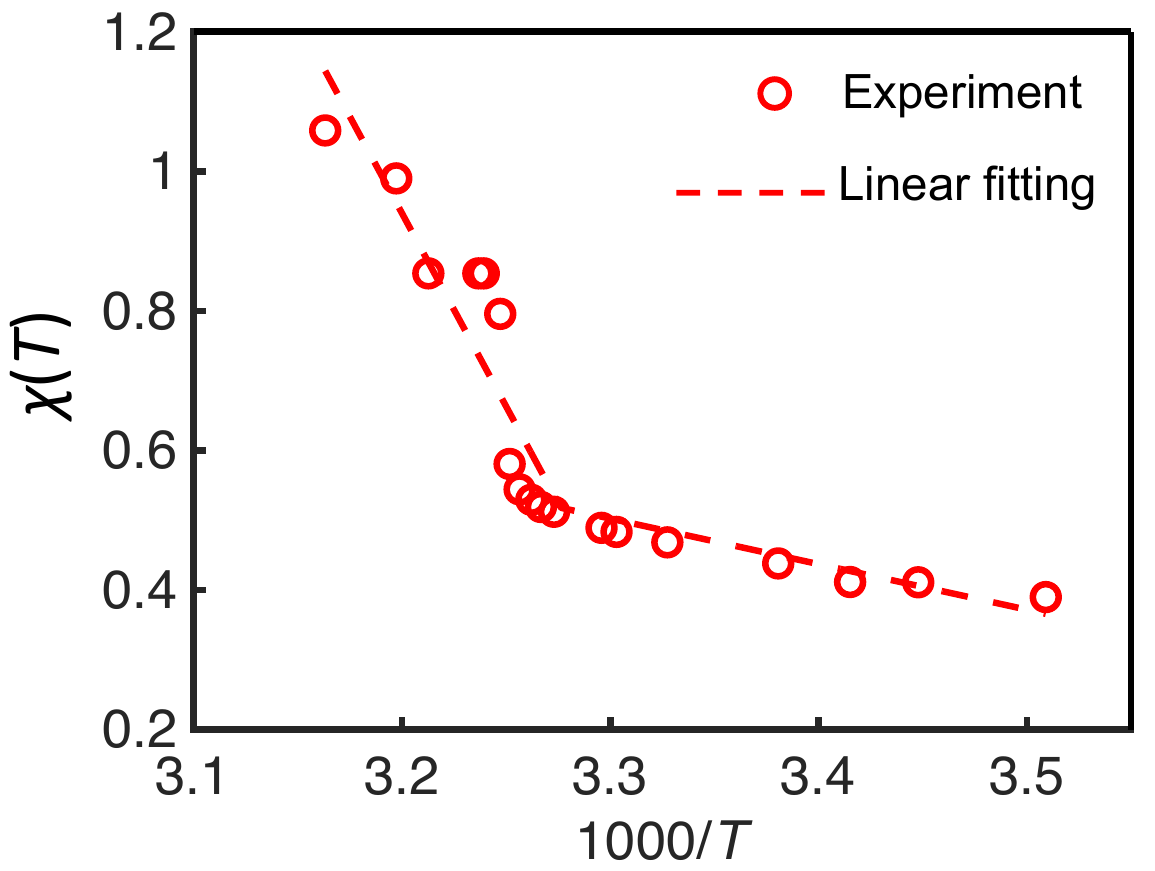}
\caption{The fitting of $\chi$ as a function of inverse temperature into two linear regimes. Temperature $T$ is in Kelvin. The experimental data is adopted from the work by Nakamoto et al. \cite{Nakamoto1996}} 
\label{figure:SM_chi_T_relation}
\end{figure}

The temperature dependence of $\chi$ is obtained from fitting the experimental data by Nakamoto et al. \cite{Nakamoto1996} The experimental data was measured from the swelling equilibrium of PNIPAM gel. As shown in Fig. \ref{figure:SM_chi_T_relation}, $\chi$ is fitted to two linear regimes as a function of inverse temperature $1/T$ ($T$ in Kelvin), given by
\begin{equation}
 \chi(T) =
    \begin{cases}
      -5.5371*\frac{1000}{T}+18.6588 & {\rm for} \ \frac{1000}{T} \leqslant 3.276 \ \ (T \geqslant 32.1 ^{\rm o}{\rm C})\\
      -0.6528*\frac{1000}{T}+2.6566 & {\rm for} \ \frac{1000}{T} > 3.276 \ \ (T < 32.1 ^{\rm o}{\rm C})
    \end{cases}   
\end{equation}

\bibliographystyle{apsrev4-2}
\bibliography{Refs}